\documentclass[sigconf]{acmart}
\usepackage{balance}

\AtBeginDocument{%
  \providecommand\BibTeX{{%
    \normalfont B\kern-0.5em{\scshape i\kern-0.25em b}\kern-0.8em\TeX}}}

\copyrightyear{2021}
\acmYear{2021}
\setcopyright{acmlicensed}\acmConference[UMAP '21]{Proceedings of the 29th ACM Conference on User Modeling, Adaptation and Personalization}{June 21--25, 2021}{Utrecht, Netherlands}
\acmBooktitle{Proceedings of the 29th ACM Conference on User Modeling, Adaptation and Personalization (UMAP '21), June 21--25, 2021, Utrecht, Netherlands}
\acmPrice{15.00}
\acmDOI{10.1145/3450613.3456835}
\acmISBN{978-1-4503-8366-0/21/06}

\begin{document}
% \fancyhf{}
%%
% \title{Explanations and User Expectations for Fairness-Aware Recommendation}
\title{Fairness and Transparency in Recommendation: The Users’ Perspective}

\author{Nasim Sonboli}
\authornote{Both authors contributed equally to this research.}
\email{nasim.sonboli@colorado.edu}
\author{Jessie J. Smith}
\authornotemark[1]
\email{jessie.smith-1@colorado.edu}
\affiliation{%
  \institution{University of Colorado Boulder}
  \city{Boulder}
  \state{Colorado}
  \country{USA}
}

\author{Florencia Cabral Berenfus}
\email{Florencia.CabralBerenfus@colorado.edu }
\affiliation{%
  \institution{University of Colorado Boulder}
  \city{Boulder}
  \state{Colorado}
  \country{USA}
  }

\author{Robin Burke}
\email{Robin.Burke@colorado.edu}
\affiliation{%
  \institution{University of Colorado Boulder}
  \city{Boulder}
  \state{Colorado}
  \country{USA}
}

\author{Casey Fiesler}
\email{casey.fiesler@colorado.edu}
\affiliation{%
  \institution{University of Colorado Boulder}
  \city{Boulder}
  \state{Colorado}
  \country{USA}
}

\renewcommand{\shortauthors}{Sonboli and Smith, et al.}

% \settopmatter{printacmref=true}
% \begin{document}
% \fancyhf{}

\begin{abstract}
Though recommender systems are defined by personalization, recent work has shown the importance of additional, beyond-accuracy objectives, such as fairness. Because users often expect their recommendations to be purely personalized, these new algorithmic objectives must be communicated transparently in a fairness-aware recommender system. While explanation has a long history in recommender systems research, there has been little work that attempts to explain systems that use a fairness objective. Even though the previous work in other branches of AI has explored the use of explanations as a tool to increase fairness, this work has not been focused on recommendation. Here, we consider user perspectives of fairness-aware recommender systems and techniques for enhancing their transparency. We describe the results of an exploratory interview study that investigates user perceptions of fairness, recommender systems, and fairness-aware objectives. We propose three features -- informed by the needs of our participants -- that could improve user understanding of and trust in fairness-aware recommender systems.
\end{abstract}




\maketitle

\section{Introduction \& Background}

In recent years, the penetration of algorithmic systems into many realms of society has raised concerns that such systems may distribute benefits and harms unfairly across individuals and groups. This is particularly evident where high stakes decisions are made, ones that have significant impact on individuals' lives and livelihoods. Personalized recommender systems are a class of algorithms that learn from past user preferences in order to predict future interests of users and provide them with suggestions tailored to their tastes. In a personalized system, the objective of the recommendation algorithm is to accurately represent users' interests. However, in recent years, non-accuracy objectives such as fairness have become more common. This is because recommender systems are increasingly being employed in higher-stakes areas such as employment and financial services. Moreover, the user -- that is, the consumer of the recommendation -- is often not the only stakeholder in a recommender system. As a result, researchers have begun to explore how to ensure that a recommender system distributes its benefits fairly within and across different stakeholder groups, such as users and item providers \cite{yao2017beyond,burke2018balanced,ekstrand2018exploring,liu2019personalized,kamishima2016model,beutel2019fairness,sonboli2020opportunistic}. In this paper, we describe findings from an exploratory interview study in which we elicited folk theories about recommender systems, as well as opinions about how fairness might be incorporated into such systems, and what forms of transparency around these algorithms are most effective for the users.
%If we have an example here, it should be for provider fairness, otherwise it can be confusing%
% For example, all else being equal, a job recommender system should not recommend executive jobs exclusively to male users and clerical jobs to female users.

%Folk Theories%
\subsection{Folk Theories to Guide Explanation Design for Fair Recommendations}
One method to collect and incorporate users' needs into a technical system is to examine their \emph{folk theories} of that system \cite{kulesza2012tell,ghori2019does}. Folk theories in Human-Computer Interaction (HCI) are defined as "intuitive, informal theories that individuals develop to explain the outcomes, effects, or consequences of technological systems" \cite{DeVito_2017}. Folk theories can help us understand how users perceive the systems that they are interacting with \cite{DeVito_2018,rader2015understanding,eslami2016first}, including uncovering inaccurate knowledge that users might have about a system in order to intervene on misguided or even harmful behavior on the platform \cite{eslami2016first, rader2015understanding, devito2018people}. These theories, accurate or not, shape the nature of user interaction and experience \cite{eslami2019user}. Therefore, it is important to broaden the education of users about the inner-workings of these systems as they lead to a better understanding of AI and web platforms  \cite{long2020ai, eslami2016first}.% Therefore, broadening the education of users leads to a better understanding of AI systems and more accurate mental models of these systems \cite{long2020ai}. 

We argue that in order to design for transparency in a fairness-aware recommender system, we must first elicit feedback from the user. Folk theories are one means to recognize where users might have inaccurate knowledge about the system and its objectives. We propose to use explanations as a way to effectively educate users about these gaps in their knowledge, which might include fairness objectives in a recommendation system. Using explanations to educate in this way can help users create more accurate mental models of these systems, which can impact their actions on a platform, and help to inform future designs \cite{Herlocker_explanation_survey}.

% We argue that in order to design for transparency in a recommender system, we must first establish effective communication between the system and the user. Folk theories allow us to recognize where users might have inaccurate knowledge about the system and its objectives, so that we may use explanations as a way to effectively educate them about these gaps in their knowledge.

%Fairness objectives in RecSys%
\subsection{Fairness Objectives in Recommendation}
Fairness in recommendation is generally assessed through two factors: representation and accuracy. Most of the proposed methods try to ensure parity in either of these factors among all user groups (e.g. women vs. men) or item groups (e.g. popular vs. unpopular). Parity is sought to avoid under-representation or over-representation of items/user subgroups, and to avoid under-estimation or over-estimation of accuracy for any item or user group \cite{yao2017beyond,Steck2018Calibrated,burke2018balanced,ekstrand2018exploring, kamishima2016model,beutel2019fairness,sonboli2020opportunistic,Sonboli2019Localized,Lin2017Fair,Ekstrand2019Fair}, as these issues could lead to discriminatory or unjust consequences for certain users.

Since recommender systems are multistakeholder systems, each stakeholder group may have different fairness needs. The key stakeholders in a recommendation system are the consumers (those who receive the recommendations, who we will also refer to as \emph{users}), the providers (those who provide items that will be recommended), and the system/platform (that hosts the recommendations)~\cite{burke2017multisided}. Fairness can be defined and sought for any of the stakeholders involved. In this paper we focus on one aspect of fairness-aware recommendation: \emph{provider fairness}. In our interviews, we frame the objective of provider fairness as a method that increases the representation of items that are consistently not being recommended.

% Transparency & Explanations&
\subsection{Transparency \& Explanations}
One of our key areas of interest in this study is the question of how to design for transparency in the context of fairness-aware algorithms. Transparency has become an essential feature of algorithmic design, especially in application areas that are socially sensitive due to (1) legal requirements for transparency such as the GDPR \cite{regulation2016regulation, goodman2017european}; (2) the need to build users' trust in a system \cite{cramer2008effects,pu2007trust,Kizilcec_too_much_info_trust}; (3) the need for error detection to help mitigate bias and discrimination in a system \cite{tintarev2007survey, phillips2020four, goetz2003matching, theodorou2017designing}; (4) helping further public adoption of new technologies \cite{kizilcec2016much, abdul2018trends, bellotti2001intelligibility, dzindolet2003role, gregor1999explanations}; and/or (5) creating an environment of accountability for the platforms that host the algorithms \cite{diakopoulos2015algorithmic, abdul2018trends}.

We believe that transparency only increases in importance for fairness-aware systems \cite{lepri2018fair}. Because of the complex and contested nature of fairness as a concept \cite{narayanan2018translation}, it is not enough to claim that a system is fair: users need to understand the specifics of the fairness objectives that have been encoded into it. It is well recognized that users of recommender systems benefit from explanations of the recommendations that they receive \cite{tintarev2007survey, recsys-xai-1, recsys-xai-2}. Because fairness-aware recommendation algorithms influence the recommendations users receive in a way that responds to fairness goals, we argue that it is particularly important that recommender systems be transparent about their fairness objectives. 

Transparency for fairness-aware recommendations is different from general transparency in recommendation algorithms (e.g., as discussed in \cite{tintarev2007survey,Rashmi2002chi}) in that the goal is not only to explain how the system works, but also to provide clarity for what its fairness goals are, the motivations for these goals, and how they might impact different stakeholders. However, explanations of fairness-aware algorithms must be done with care; the \emph{ways} in which fair algorithms are explained to users have been proven to heavily influence their trust, adoption, and understanding of these systems \cite{perceived-fairness, explaining-injustice}. Since explanations in a fairness-aware system impact users' trust of and actions within the system, it is especially important to ensure that these explanations are created effectively and carefully, with the users' needs in mind.

While explanation has a long history in recommender systems research \cite{tintarev2007survey, recsys-xai-1, recsys-xai-2}, there has been little work that addresses explaining systems that involve a fairness objective. Similarly, while previous work in other branches of AI has explored the use of explanations as a tool to increase fairness (for example, by reducing unwanted bias or discrimination) \cite{arrieta2020explainable, wang2019designing, doshivelez2017rigorous}, this work has not focused on recommendation specifically.

% To fill this gap in knowledge, our contributions in this paper are threefold. First, we suggest that folk theories are an effective way to begin designing explanations in fairness-aware recommender systems due to their ability to discern gaps in user knowledge about the system. Second, we propose that explanations of fairness-aware systems should include (1) definitions of fairness goals; and (2) an organization's motivations for selecting fairness as their algorithm's objective. Finally, we argue that explanations of fairness-aware systems should also be used for \emph{education} about issues of algorithmic unfairness and as a tool for transparency, rather than a weapon for manipulation.

% One of the main takeaways from this research was the confirmation that folk theories do act as useful method to uncover the gaps and inaccuracies that exist in user knowledge of a system. Drawing from this knowledge, o
This paper contributes findings about participants' folk theories of recommendations and fairness, as well as their opinions about transparency and explanation. Based on these findings, we take the position that explanations ought to be used as a means to educate users about algorithmic objectives, especially in fairness-aware systems, and suggest three features that should be included in effective explanations of fairness-aware recommender systems.

% 1 -- one of our biggest takeaways is that folk theories can be utilized as a method to recognize the gaps that exist in user knolwedge of a system.
% 2 -- utilizing this understanding, our contributions in this paper reflect that we believe that explanations of fairness-aware recommender systems should be used for three purposes: (1) to EDUCATE users about fairness in recommender systems and the problems that accuracy-based systems might create, (2) to explain the definition of fairness that has been chosen for the objective of an algorithm, (3) to explain the motivation behind this choice, and its impact on the end user.

\section{Methods}

In this work, we use interviews as a way to explore the stories and experiences of a sample of users of recommender systems. Though some previous work has stated that user input for algorithmic design may not be helpful for engineers, due to users' potential lack of knowledge of the system \cite{Saxena2019_fairdefs}, we argue that it is precisely this lack of knowledge that can be beneficial for the design and deployment of \textit{explanations} in systems. Understanding what the users do \emph{not} know about a system can help the engineer pinpoint where users might need more education about the platforms they are interacting with, and how an explanation can meet those needs.

As an initial exploratory study, we recruited a convenience sample of 30 undergraduate and graduate students from a large state university in the United States, all of whom had prior exposure to recommender systems (e.g., on e-commerce or streaming platforms such as Amazon or Netflix). Our participants included 18 women and 12 men, with an age range of 18 to 32. 15 participants were studying Computer Science, Information Science, or related fields, and 15 were pursuing other majors. Participants were recruited via  social media postings, a university announcement board, and advertisements in classes, and were compensated \$10 for their time; this study was approved by our institution's IRB. Though our sample of young internet users provides a good foundation for understanding opinions and concerns that ordinary recommendation consumers might have about fairness, we recognize that it is limited in scope and we do not suggest that our findings are generalizable to a broader population.

We conducted face-to-face, semi-structured interviews with participants that began with a discussion of their past experiences with recommender systems and their folk theories of recommendation algorithms and fairness objectives. We asked them to explain their current knowledge about how these systems work, and what fairness meant to them in this context. For the latter half of the interviews, we used the Kiva crowd-sourced microlending platform (explained in more detail below) as a case study to explore the ways that explanations can help or harm the participant's understanding of a fairness-aware system.

Following interview transcription, researchers conducted thematic analysis \cite{braun2006using}. Three independent analysts conducted open-coding on a subset of the interviews using MAXQDA software, and then met to confer on, synthesize, and finalize a set of themes that best captured the insights gathered from our participants. We conducted a detailed analysis of each theme, resulting in the overall story that our data tells, as described in detail in the Results and Discussion sections.

\subsection{Kiva as a Case Study}
% and why Kiva is a good case study for fairness in recsys
Kiva is an organization seeking to enhance financial inclusion in the world through microlending. Their platform provides a space for those who are seeking loans (borrowers) to get funded by those who are seeking to supply capital (lenders). Kiva's mission emphasizes equitable access to capital for all borrowers, who are individuals without access to traditional forms of capital, to improve their living situations. If Kiva implemented a recommender system, the providers of the recommended content would be the borrowers, and the consumers of the recommendations would be the lenders \cite{Choo_understanding_kiva}. Naturally, a recommender system in this context raises an important fairness concern. If borrowers are recommended inequitably, the system could be contributing to an inequitable pattern of resource distribution on its platform. 
We used Kiva as a case study both to have a concrete example to ensure consistency across participants, and because Kiva is a real-world example that contains fairness concerns. In our interviews, we asked the participants to consider what helpful explanations might look like in Kiva's platform if fairness-aware recommendation were implemented.

\section{Results}
In this section, we describe the main findings from our exploratory interviews with users of recommender systems. We begin by explaining some of the gaps in participants' knowledge about recommender systems and fairness objectives, as gathered from participant folk theories. Then, we explore the ways in which participants indicated that explanation design of a fairness-aware system could help or harm their understanding and trust of the system. Participant quotes are indicated by anonymized participant numbers; participants 1-15 were studying CS or adjacent fields, and 16-30 were in non-technical fields.

\subsection{Folk Theories of Recommender Systems}
Some participants expressed that the “black box” nature of recommendations made it difficult for them to learn how a recommendation system might create personalized recommendations for them. For example, as P5 said, "recommender systems in most cases are pretty unknown to the user."

However, in general, participants had somewhat of an understanding for how recommender systems worked -- particularly in terms of how much data was being collected on them. Some participants' folk theories aligned well with  accuracy-based recommendation algorithms, such as collaborative filtering algorithms like User-KNN or Item-KNN. 

\begin{quote}
“So if like a lot of people buy, you know, a hammock, then if a large subset of those people who bought a hammock, also bought a sleeping bag, it might recommend the sleeping bag to you.” -- P21
\end{quote}
% \begin{quote}
% “A lot of it's just gathering data about you and then making recommendations and suggestions based on projected trends that they see.” -- P11 
% \end{quote}

% When asked about the goals of an organization that utilizes recommendations, many users expressed that they thought the main goal was money, even if it was obtained at the expense of the consumer's experience on the platform, as expressed by P5.

% \begin{quote}
% “I mean I think they want to have a good system so that I continue spending money, but I feel like their top priority is money.” -- P5
% \end{quote}

\subsection{Folk Theories of Fairness Objectives}
When asked about how recommendations could be unfair to users, very few participants indicated that they had ever thought of provider fairness for recommendations. This raised a concern that there might be a lack of communication between the system and the user when it comes to issues of provider fairness in recommendation. 

After a short discussion about fairness objectives and the impact that a recommender system might have on the providers, many participants indicated that they thought provider fairness was important to them in a recommender system, as expressed by P18.

\begin{quote}
“I don't think I've thought about it as much from a seller's perspective, but I can see like, you know, these platforms are made for more than just [users]. They're made for [providers] as well. So like people who sell things on Amazon or make music and put it on Spotify, it's like, yeah, you're probably inherently at a disadvantage to people who are already big or are already in favor of the algorithm.” -- P18
\end{quote}

% Overall, there was some inconsistency in the ways that participants theorized about the functionality of recommender systems in terms of fairness. Some participants were very knowledgeable about these topics, while others had little-to-no understanding of how they functioned at all. 

However, the majority of participants were still not entirely sure how fairness goals might impact their recommendations. Thus, we began to explore different ways that a recommendation platform could explain this impact to users in an educational and empowering way.

\subsection{Explaining Fairness Goals to Users}
Throughout the interviews, most participants indicated that they would want to see the fairness goals of an organization described to the users in some way, either through a short explanation or an entire page, as described by P8. 
\begin{quote}
“[Organizations] should have [fairness goals] somewhere I could find it like the little ‘about us’,  like ‘learn more about our corporation’ tab where they would explain ‘these are our moral values, these are what we prioritize’.” -- P8
\end{quote}

One participant thought that fairness goals were best incorporated into the UX of the platform, as long as the language did not unintentionally manipulate users. \begin{quote}
“Or like work [fairness goals] into their platform somehow. Like how Spotify does the ‘New Music Fridays’... I think that there are opportunities to weave that into places on a platform but not necessarily blanket it across so the user doesn't feel like they have choice.” -- P29 
\end{quote}

This raises an important concern that is not new in the field of explanation: the concern that explanations, if not designed carefully, could be used as a tool to manipulate users.

\subsection{Fairness as a User Choice}

One outcome of fairness-aware recommendation is to "nudge" users into choosing items that they might otherwise not. The intention of this kind of system is to encourage the user to choose the items that meet the fairness objective of the system itself, and it is precisely this intention that might underlie some participants' concerns with being "manipulated". In fairness-aware recommender systems, \emph{coercing} users into choosing a "fair" recommendation -- which might not be 'fair' for everyone -- could be misleading. This concern was expressed amongst our participants, particularly because fairness definitions are often disputed, and what might be considered \emph{fair} for some, could seem \emph{unfair} for others \cite{smith2020exploring}. In order to remedy these concerns, several of our participants indicated that they would prefer to be informed about the existence of fairness-aware recommendations so that the user could make the choice of fairness or personalization for themselves, as expressed by P22.

\begin{quote}
“[Fairness-aware recommendation] is manipulative in some sort of way. I think the best thing that they can do would be to give a short explanation that they changed [the personalized recommendation algorithm] and then kind of show the whole list and not point out any others in the list... allowing an unbiased choice from the viewer.” -- P22
\end{quote}

% Several participants expressed that framing recommendation decisions around the consumer's experience was very important for them to maintain trust in the system. As explained previously, consumers need to know that their experience is still important to the organization -- regardless of fairness goals. 
% \begin{quote}
% “If the company frames these [explanations] as ‘we want to help the customer be satisfied in some way’, then people will be fine with it. If it's somehow presented in a way of like, ‘we want to push something on you that you wouldn't normally do’, then I don't think people are going to be happy.” -- P15
% \end{quote}

If manipulation was a concern amongst participants, then how should explanations be designed to mitigate that concern? Further, how might explanations instead be designed to foster trust and communication between a system and its users? In the next section, we propose that these concerns can be alleviated through explanations if they are used effectively as a tool for education.

\subsection{Explanations as Education}

One prominent theme that emerged throughout the interviews was that \textit{transparency as a means for education was essential}. Whether an algorithm uses fairness objectives or not, participants expressed that they needed to be educated about why they were being recommended the items that they were. This point was succinctly summed up by P21 who stated: “the more transparency, the better.”

Many participants indicated that better design practices could help promote fair treatment for providers and relay this fair treatment back to the recommendation consumers. In order to ground participants in a specific platform and discuss concrete explanation designs, we used Kiva as a case study. Specifically, we asked the participants how they thought transparency could be designed when fairness goals were incorporated in a system like Kiva. We gathered feedback about the benefits and flaws of explanation designs that were minimal and global, versus detailed and specific.

Several participants were in favor of including one global explanation for all fairness-aware recommendations. Some indicated that a single explanation could still leave room for the system to include more in-depth explanations about a specific group of providers that the system is optimizing for fairness. For example, many participants expressed that they would like to know more about why Kiva borrowers from a specific geographic area are underfunded on the platform, and how choosing to lend to people from that area could have a positive social impact.
\begin{quote}
“I like [longer explanations], where you can scroll underneath and get like a broader perspective of what's going on in the whole region, I would like that.” -- P5
\end{quote}

% \subsubsection{\textbf{Transparency With Detailed Explanations}}
% %\linebreak
Another alternative was to provide a separate explanation for every individual recommendation, on the recommended item itself. Many participants expressed that if explanations were specific to each recommendation, the manner in which they were formatted played a big role in their utility to the consumer. P30 described that for explanations to be effective and informative, the \textit{“algorithms should be explained to you in a simple sentence structure... short, concise and in digestible pieces”}. They added that explanations should be accessible and inclusive for all audiences, regardless of their level of understanding of algorithms.

% Participants expressed that this design would only work if each explanation was specific to that recommended item, or else it would get redundant to the consumer. 
% \begin{quote}
% “If all of my recommendations are the same, then like, I don't want to be seeing the same paragraph over and over.” -- P6
% \end{quote}

% \begin{quote}
% “Describing how the algorithm works to a third grader, like if you're explaining to a third grader what this algorithm does, there should be a separate section for that.” -- P30
% \end{quote}

In summary, we found that when it comes to transparency for fairness-aware recommendations, design decisions matter. Transparency should give users greater agency to understand how provider fairness might make their recommendations different than personalization. Further, when recommendation fairness is explained to the consumer, it should be explained in a way that is accessible, understandable, specific, and concise.

\section{Discussion \& Conclusions}
In this study, we explored what users of recommender systems understand about personalized and fairness-aware recommendations. Through semi-structured interviews, we used folk theories as an effective method for pinpointing gaps and inaccuracies in user knowledge of a system. This formed our basis for understanding that transparency in recommender systems ought to serve as a means of education for users, especially in a fairness-aware system where the objectives are often unknown or misunderstood by the user. Taking into account the feedback provided by the interview participants, we provide the following suggestions for features that should be included in effective explanation design for fairness-aware systems.

\begin{enumerate}
    \item \emph{Explanations should define the system's fairness objective for users.} For example, an explanation could educate the user about the impact that a fairness-aware recommendation might have on the item providers.
    \item \emph{Explanations should not nudge/manipulate users into making a decision, even if the goal is fairness.} For example, an explanation could educate users about the existence of a fairness-aware recommendation while still allowing the user to decide if they prefer personalization instead.
    \item \emph{Explanations should disclose the motivation for using fairness as a system objective.} For example, an explanation could educate users about the fairness concerns of the system, or the values of the organization and why they chose fairness as an objective.
\end{enumerate}

These three features were compiled based on the specific gaps that our participants had in their knowledge about fairness-aware recommender systems. Specifically, the folk theories that participants shared on these systems led us to believe that there might exist a major gap in communication towards users about recommender systems' objectives. These features also summarize participants' most frequent desires for what they wished explanations could teach them about the system in order for them to understand and trust it more. It is also interesting to note that whenever participants were educated about provider fairness concerns in recommendations, they generally showed interest in the option of using a system that had fairness as an objective. 

However, we note some limitations in this study's exploration of participants' opinions around fairness, transparency and explanation. Firstly, presenting examples of provider fairness in a face-to-face interview setting could have resulted in acquiescence response bias, or a tendency for participants to frame their views in ways that they believed was expected from or valued by the interviewer \cite{liu2017acquiescent}. Moreover, it is possible that social desirability response bias could have influenced participants’ expressed interest in provider fairness, when discussing Kiva as a nonprofit organization seeking to increase financial access globally \cite{randall1991socialdesir,vandeMortel2008socialdesir}. It may also be the case that recommender systems that do not carry such high stakes for providers (e.g., a food delivery recommender system) might attract different levels of interest from users when it comes to provider fairness. Thus, we suggest that future work could further investigate similar questions by way of anonymous self-reports, or by assessing participant behavior in an experimental design.

With this work, we encourage organizations that are trying to incorporate fairness into the design of their recommender systems to seek input from their users when designing for transparency. 
% TODO: suggest future work by the research community also
If transparency design fails to consider the knowledge (or lack thereof) of the users, the platform runs the risk of losing user trust. Moreover, if fairness goals are explained to users in vague or ineffective ways, this can cause confusion or lead to misunderstanding of the goals of the system. In turn, when fairness goals are explained to the users effectively, they can gain agency, and be left feeling more empowered and knowledgeable about the platform and their impact while using it. Ultimately, we argue that in order to gain users' trust in and adoption of a fairness-aware recommender system, collaboration is \emph{necessary}. Through effective communication, feedback, and transparency, recommender systems can contribute to a more empowering, fair, and trustworthy digital future.

\section{Acknowledgements}
This work was supported by the National Science Foundation under Grant No. 1911025.

\vspace{0.25 in}

\bibliographystyle{ACM-Reference-Format}
\balance
\bibliography{main}

\end{document}